\documentclass[11pt]{amsart}

\def\cal{\mathcal}

 \date{}
 
\begin{document}

% \maketitle
\def\abstractname {Abstract}
\newtheorem {lemma} {\bf Lemma}
\newtheorem {theor} {\bf Theorem}
\newtheorem {cor} {\bf Corollary}
\newtheorem {z} {\bf Proposition}

\author{Krassimir Yankov Iordjev, Dimiter Stoichkov Kovachev}

\address{
Department of computer science \\ South-West University ``Neofit Rilski``\\Faculty of Mathematics and Natural Sciences\\
2700 Blagoevgrad \\ Bulgaria. } \email{iordjev@swu.bg \;- Krassimir Yankov Iordjev} \email{dkovach@abv.bg \;- Dimiter Stoichkov Kovachev }

%\address{Department of computer science\\
%South-West University ``Neofit Rilski``, Blagoevgrad \\
%Krassimir Yankov Iordjev,  e-mail: iordjev@swu.bg \\
%Dimiter Stoichkov Kovachev,  e-mail: dkovach@abv.bg }

\title[On finding a particular class of combinatorial identities]
{On finding a particular class of combinatorial identities}

%\begin{document}
%\baselineskip 22pt

 \begin{abstract}
 In this paper, a class of combinatorial identities is proved. A method is used which is based on the following rule: counting elements of a given set in two ways and making equal the obtained results. This rule is known as ``counting in two ways". The principle of inclusion and exclusion is used for obtaining a class of $(0,1)-$matrices.

\end{abstract}

\maketitle

%\section{  Introduction, Definitions and Notation}
A modification of the method of ``counting in two ways" (\cite{1}, p.2) is to obtain a general formula about the number of elements of a set, after that to consider some of the subsets of this set, and to find elements of these subsets, on the one hand, by using the general formula, and on the other hand, by using specific properties of the subsets.

We will demonstrate this method by considering and proving the following identities:

\begin{equation}\label{n=k,p=1}
\sum_{s=0}^{n-1} (-1)^s {n\choose s} (n-s)^n =n!
\end{equation}

\begin{equation}\label{n=2p,k=2}
\sum_{s=0}^{p} (-1)^s {{2p}\choose s} {{2p-s}\choose p}^2
=\frac{(2p)!}{(p!)^2}
\end{equation}

\begin{equation}\label{n>pk}
\sum_{s=0}^{n-p} (-1)^s {{n}\choose s} {{n-s}\choose p}^k =0,
\; {\mbox{where}} \; n>pk
\end{equation}

\begin{equation}\label{n=kp}
\sum_{s=0}^{p(k-1)} (-1)^s {{kp}\choose s} {{kp-s}\choose p}^k
=\frac{(kp)!}{(p!)^k}
\end{equation}

\begin{equation}\label{p=1}
\sum_{s=0}^{n-1} (-1)^s {{n}\choose s} (n-s)^k = \sum_{\scriptsize
\begin{array}{c} (t_1 ,t_2 , \ldots , t_n ),\; t_i \geq 1
\\t_1 + t_2 + \cdots +t_n =k
\end{array}} \frac{k!}{t_1 ! t_2 ! \ldots t_n !} ,\; {\mbox{if}}\ n\le k
\end{equation}

What these identities have in common is the left-hand side that can be written by the expression:

\begin{equation}\label{rnkp}
{\cal{R}} (n\times k,p) =\sum_{s=0}^{n-p} (-1)^s {n\choose s}
{n-s\choose p}^k
\end{equation}

and (\ref{n=k,p=1}) is obtained with $n=k$ and $p=1$;
(\ref{n=2p,k=2}) with $n=2p$ and $k=2$; (\ref{n>pk}) with $n>pk$;
(\ref{n=kp}) with $n=kp$; (\ref{p=1}) with $p=1$.

It remains to give some of the possible combinatorial interpretations of the expression (\ref{rnkp}). Boolean (binary, or (0,1)-matrix) is a matrix whose entries are equal to 0 or 1. We will show that ${\cal{R}} (n\times k,p)$ gives the number of all $n\times k$ (composed of $n$ rows and $k$ columns) Boolean matrices, such that in each column they have exactly $p$ ones, and in each row these matrices have at least 1 one. Necessary and sufficient condition for existence of such matrices is $1\le p\le n\le kp,$ as in the special case when $n=kp$, in each row we have exactly 1 one, and when $n>kp$, such matrices do not exist, that is, their number is equal to zero (see Proposition 3). Indeed, it is easy to see that the number $r(n\times k,p)$ of all $n\times k $ Boolean matrices, that have exactly $p$ ones in each column, is equal to

\begin{equation} \label{r}
r(n\times k,p)={n \choose p}^k
\end{equation}

From the set of all $n\times k $ Boolean matrices, that have exactly $p$ ones in each column, we have to remove matrices that have at least one row of zeroes, and to determine the number of the remaining matrices. We will do this by using the principle of inclusion and exclusion. Recall this known principle in the following form:

{\bf Inclusion and exclusion principle}: Let $M$ be a finite set and let $P=\{ p_1 ,p_2 ,\ldots ,p_m\}$ be the set of properties that can be possessed by the elements of $M$. Denote by $N(p_{i_1} ,p_{i_2} ,\ldots ,p_{i_s} )$ the number of elements of $M$ that possess the properties $p_{i_1},p_{i_2} ,\ldots ,p_{i_s}$. Then the number $N(\emptyset )$ of elements of $M$ that do not possess any of the properties of $P$ is given by the formula:

$$N(\emptyset )=|M|+\sum_{s=1}^m (-1)^s \sum_{1\leq i_1 <i_2 <\ldots <i_s} N(p_{i_1} ,p_{i_2} ,\ldots ,p_{i_s} ) =$$

$$  =|M|-\sum_{j=1}^m N(p_j )+\sum_{1\leq i_1 <i_2}^m N(p_{i_1} ,p_{i_2}) - \sum_{1\leq i_1 <i_2  <i_3}^m N(p_{i_1} ,p_{i_2} ,p_{i_3} ) +\cdots +$$

$$+(-1)^s \sum_{1\leq i_1 <\cdots <i_s}^m N(p_{i_1} ,\ldots ,p_{i_s} )+\cdots + (-1)^m N(p_1 , p_2 ,\ldots , p_m ).$$

In particular, in our problem let $M$ be the set of all $n\times k$ Boolean matrices that have in each column exactly  $p$ $(1\le p\le n)$ ones, and the property $p_i$ is possessed by these matrices belonging to the set $M$ whose $i$-th row, $1\le i\le n$, consists of zeroes. Since $M$ cannot contain matrices with more than $n-p$ rows of zeroes, then  $N(p_{i_1} ,p_{i_2} ,\ldots ,p_{i_s})  = 0$ when $s>n-p.$
Then obviously $|M|=r(n\times k,p)$, and for each set of properties $ \{p_{i_1} ,p_{i_2} ,\ldots ,p_{i_s}\}$ we have $N(p_{i_1} ,p_{i_2} ,\ldots ,p_{i_s})  = r((n-s)\times k,p)$, $\ s=1,\ 2,\ldots,\ n-p$.

Since $s$ of $n$ properties $ p_1 ,p_2 ,\ldots ,p_n $ can be chosen in $\displaystyle {n\choose s}$ ways, then
  $$\displaystyle \sum_{1\leq i_1 < i_2 < \cdots <i_s}^n N(p_{i_1}, p_{i_2} , \ldots ,p_{i_s} )=
   \displaystyle {n\choose s}r((n-s)\times k,p)$$.

From (\ref{r}), applying the inclusion and exclusion principle, we get:

$${\cal{R}} (n\times k,p)=|M|+\sum_{s=1}^{n-p} (-1)^s {n\choose s} {n-s\choose p}^k =$$
$$ {n \choose p}^k + \sum_{s=1}^{n-p} (-1)^s {n\choose s} {n-s\choose p}^k =\sum_{s=0}^{n-p} (-1)^s {n\choose s} {n-s\choose p}^k ,$$
quod erat demonstrandum.

In order to prove the identities (\ref{n=k,p=1}) - (\ref{p=1}), it remains to prove following propositions:

\begin{z}
Prove that the number of all quadratic Boolean matrices with $n$ rows and $n$ columns, having in each row and in each column exactly 1 one, is equal to $n!$.
\end{z}

\begin{z}
Prove that the number of all Boolean matrices with $2p$ rows and two columns, having in each row 1 one and 1 zero and the same number of ones in each column, is equal to
$\displaystyle
\frac{(2p)!}{(p!)^2}$.
\end{z}

\begin{z}
Prove that in a Boolean matrix with $n$ rows and $k$ columns, such that in each column there is exactly $p$ ones and $n>kp$, there exists at least one row of zeroes.
\end{z}

\begin{z}
If ${\cal{B}}$ is the set of all Boolean matrices with $n$ rows and $k$ columns, such that in each column there is exactly $p$ ones, $n=kp$ and there are no rows of zeroes, prove that
$|{\cal{B}}|=\displaystyle \frac{(kp)!}{(p!)^k}$.
\end{z}

\begin{z}
If ${\cal{E}}$ is the set of all Boolean matrices with $n$ rows and $k$ columns, $n\le k$, that have in each column exactly 1 one and there are no rows of zeroes, prove that
$$|{\cal{E}}|=  \sum_{\scriptsize
\begin{array}{c} (t_1 ,t_2 , \ldots , t_n ),\; t_i \geq 1
\\t_1 + t_2 + \cdots +t_n =k
\end{array}} \frac{k!}{t_1 ! t_2 ! \ldots t_n !} .$$
\end{z}

Proofs of Proposition \ref{n=kp} and identity (\ref{n=kp}):
\begin{proof}
 From $n=kp$ it follows that in each row of a matrix of the set ${\cal{B}}$ there is exactly 1 one and $|{\cal{B}}|={{\cal{R}} (n \times k,p)}={\cal{R}}(kp \times k,p).$ Let $ C,\ C\in{\cal{B}}$ be an arbitrary matrix and the ordered $n-$tuple $(c_{1t_1},\ c_{2t_2},\ldots,\ c_{nt_n})$ consists of the nonzero elements of this matrix. With these elements, we associate numbers of their columns, respectively, namely the ordered $n-$tuple $(t_1,\ t_2,\ldots,\ t_n).$ Conversely, with the $s-$th element $t_s$, we can associate the nonzero element of row $s$ of matrix $C$, namely $c_{st_s}.$ If $(d_{11},\ d_{23},\ d_{31},\ d_{43},\ d_{52},\ d_{62})$ are the nonzero elements of matrix $ D,\ D\in{\cal{B}},\ k=3,\ p=2,$ we associate the ordered 6-tuple $(1, 3, 1, 3, 2, 2)$ with them. Since Boolean matrices are uniquely determined by their nonzero entries, then the number of matrices of the set ${\cal{B}}$ is equal to the number of different permutations of $(t_1,\ t_2,\ldots,\ t_n)$, that is, equal to
$(1,\ldots,\; 1,\; 2,\ldots,\; 2,\ldots,\; k, \ldots,\; k)$, where each number of a column (columns are $k$ in number) is repeated exactly $p$ times, where $p$ is the number of ones in a column. The number of these permutations with repetitions (\cite{1} - the number of permutations of length $n$, composed by $s$ different elements, repeated $q_1,\ q_2,\ldots ,\ q_s$ times, respectively, is given by the expression\\

$ \displaystyle
\frac{n!}{{q_1}!{q_2}! \ldots {q_s}!}$, where $n=q_1+q_2+\ldots +q_s$)
is equal to
$$ \displaystyle
\frac{n!}{(p!)^k}=\displaystyle
\frac{(kp)!}{(p!)^k}=|{\cal{B}}|={\cal{R}}(kp \times k,p).$$
The proof is completed. \\
\end{proof}

From Proposition \ref{n=kp}, with the special case when $p=1, (k=2)$, we obtain Proposition \ref{n=k,p=1} (Proposition \ref{n=2p,k=2}).\\

Proofs of Proposition \ref{p=1} and identity (\ref{p=1}):
\begin{proof}
Each Boolean matrix of the set ${\cal{E}}$ contains $k$ ones - exactly 1 one in the column and at least 1 one in each row. Therefore
$|{\cal{E}}|= {{\cal{R}} (n \times k,1)}.$
If $ C,\ C\in{\cal{E}}$ is an arbitrary matrix, the ordered $k-$tuple $(c_{r_1 1},\ c_{r_2 2},\ldots,\ c_{r_k k})$ consists of its nonzero elements, then let $t_i$ be the number of ones in the $i-$th row, where $t_i \geq 1,\ i=1,\ 2,\ldots,\ n$ and $t_1 + t_2 +\ldots + t_n =k $.
With matrix  $C$, we can uniquely associate the ordered $k-$tuple  $(c_{r_1 1},\ c_{r_2 2},\ldots,\ c_{r_k k})$ or the ordered $k-$tuple $(r_1,\ r_2,\ldots,\ r_k)$. The number of Boolean matrices of ${\cal{E}}$, that have $t_i$ ones in the $i-$th row, is equal to the permutations $(r_1,\ r_2,\ldots,\ r_k)$, that is, to the number of permutations of the $k-$tuple $(1,\ldots ,\; 1,\ldots,\; 2, \ldots ,\; 2, \ldots ,\; n, \ldots,\;  n)$, where the number $i, \ i=1,\; 2,\ldots,\; n$ is repeated $t_i$ times.
This number is equal to $ \displaystyle \frac{k!}{{t_1}!{t_2}! \ldots {t_n}!}$, and concerning the number of matrices of the set ${\cal{E}}$ we get
$$|{\cal{E}}|= \displaystyle \sum_{\scriptsize
\begin{array}{c} (t_1 ,t_2 , \ldots , t_n ),\; t_i \geq 1
\\t_1 + t_2 + \cdots +t_n =k
\end{array}} \frac{k!}{t_1 ! t_2 ! \ldots t_n !},$$
where the sums are taken over all possible expansions of number $k$ into $n$ nonzero terms. \\
\end{proof}


\begin{thebibliography}{5}

\bibitem{1} M. Eigner, Combinatorial Theory, Springer, 1997.

\end{thebibliography}
\end{document}